\documentclass[twocolumn,aps,,reprint]{revtex4-1}
\usepackage{amsmath,amsfonts,amssymb}
\usepackage[english]{babel} 
\usepackage{hyperref}
\usepackage[T1]{fontenc}
\usepackage{color}
\usepackage{float}
\usepackage{verbatim}
\usepackage{graphicx}
\usepackage{bm}
\usepackage{mathtools}

\newcommand{\incrementD}[1]{ \langle[X(t + T) - X(T)]^2 \rangle  }
\usepackage{graphicx}% Include figure files
\usepackage{dcolumn}% Align table columns on decimal point
\usepackage{bm}% bold math
\usepackage[english]{babel}
\usepackage{amsmath}

\begin{document}

\preprint{APS/123-QED}

\title{Self-interacting random walks : aging, exploration and first-passage times }% Force line breaks with \\

\author{A. Barbier--Chebbah}
\affiliation{ 
Laboratoire de Physique Th\'eorique de la Mati\`ere Condens\'ee, CNRS, UPMC, 4 Place Jussieu, 75005 Paris, France %\\This line break forced with \textbackslash\textbackslash
}%
\author{O. B\'enichou}\affiliation{ 
Laboratoire de Physique Th\'eorique de la Mati\`ere Condens\'ee, CNRS, UPMC, 4 Place Jussieu, 75005 Paris, France %\\This line break forced with \textbackslash\textbackslash
}%
%\affiliation{%


\author{R. Voituriez}
\affiliation{ 
Laboratoire Jean Perrin, CNRS, UPMC, 4 Place Jussieu, 75005 Paris, France %\\This line break forced with \textbackslash\textbackslash
}
\affiliation{ 
Laboratoire de Physique Th\'eorique de la Mati\`ere Condens\'ee, CNRS, UPMC, 4 Place Jussieu, 75005 Paris, France %\\This line break forced with \textbackslash\textbackslash
}
%\affiliation{
%\affiliation{
 %Third institution, the second for Charlie Author
%\affiliation{%


\date{\today}% It is always \today, today,

\begin{abstract}
Self-interacting random walks are endowed with long range memory effects that emerge from the interaction of the random walker at time $t$ with the territory that it has visited at earlier times  $t'<t$. This class of  non Markovian random walks has  applications in a broad range of examples, ranging from insects to living cells,  where a random walker modifies locally its environment -- leaving behind  footprints along its path, and in turn responds to its own footprints. Because of their inherent non Markovian nature, the exploration properties of self-interacting random walks have remained elusive. Here we show that long range memory effects can have deep consequences on the dynamics of generic self-interacting random walks ; they can induce  aging  and non trivial persistence and transience exponents, which we determine quantitatively, in both infinite and confined geometries. Based on this  analysis, we quantify the search kinetics of self-interacting random walkers and show that the distribution of the first-passage time (FPT)  to a target site in a confined domain takes   universal scaling forms  in the large domain size limit, which we characterize quantitatively.  We argue that memory abilities induced by  attractive self-interactions provide a decisive advantage for local space exploration, while repulsive self-interactions can significantly accelerate the global exploration of large domains.

\end{abstract}

\maketitle

%\tableofcontents

%\section{Content}

\section{Introduction}

%\subsection{Definitions : self-interacting random walks}

%---importance of random walks that interact with their own path. Relevance in ecology (ants, cells...).

Random walk theory provides a natural framework  to model transport processes at all scales. Beyond the historical examples provided by particle transport in simple fluids at the molecular and supramolecular scales \cite{Kampen:1992,Hughes:1995,sokolovbook}, it has also proved more recently to  powerfully describe  the dynamics of more complex, passive or active, larger scale  systems -- ranging from polymers, molecular motors or self-propelled colloids to cells or animals, whose dynamics take place in potentially complex environments \cite{Bouchaud:1990b,D.Ben-Avraham:2000,R.Metzler:2000,Burioni:2005,Romanczuk:2012fk,Metzler:2014dz,Bechinger:2016wt}.   In the latter  case,  the coupling of the internal degrees of freedom of the random walker to those of the environment generically leads to complex correlations and require a non Markovian description of the evolution over time of the position $X(t)$  of the random walker. Taking into account such memory effects  remains a theoretical challenge  even if several examples of model systems have been analyzed \cite{Masoliver:1986mz,Bicout:2000,Schutz:2004ul,GuerinT.:2012fk,Boyer:2014uz,falcon-cortes_2017,Guerin:2016qf,Guerin:2016qf,rieger}.

In this paper we focus on a broad class of such non Markovian random walkers, where memory effects emerge from the interaction of the random walker at time $t$ with the territory that it has visited at earlier times  $t'<t$  \cite{amit_asymptotic_1983, peliti_random_1987, ottinger_generalised_1985,toth_self-interacting_2001, sapozhnikov_self-attracting_1994, pemantle_survey_2007,foster_reinforced_2009, ordemann_structural_2001, davis_reinforced_1990}. This class of self-interacting random walks has clear applications in a broad range of examples where a random walker modifies locally its environment -- leaving behind  footprints along its path, and in turn responds to its own footprints \cite{Kranz:2016vy,Kranz:2019uw}.  Such behaviours have been reported  for   ants depositing pheromones along their path \cite{Dussutour:2004uv}, larger territorial animals \cite{Giuggioli:2011vu}, and have been identified quantitatively in the case of living cells  that chemically modify and remodel the extra-cellular matrix  \cite{dAlessandro:2021wx,Flyvbjerg:2021to}.

\begin{figure}[h!]
    %\centering
    \includegraphics[scale=0.27]{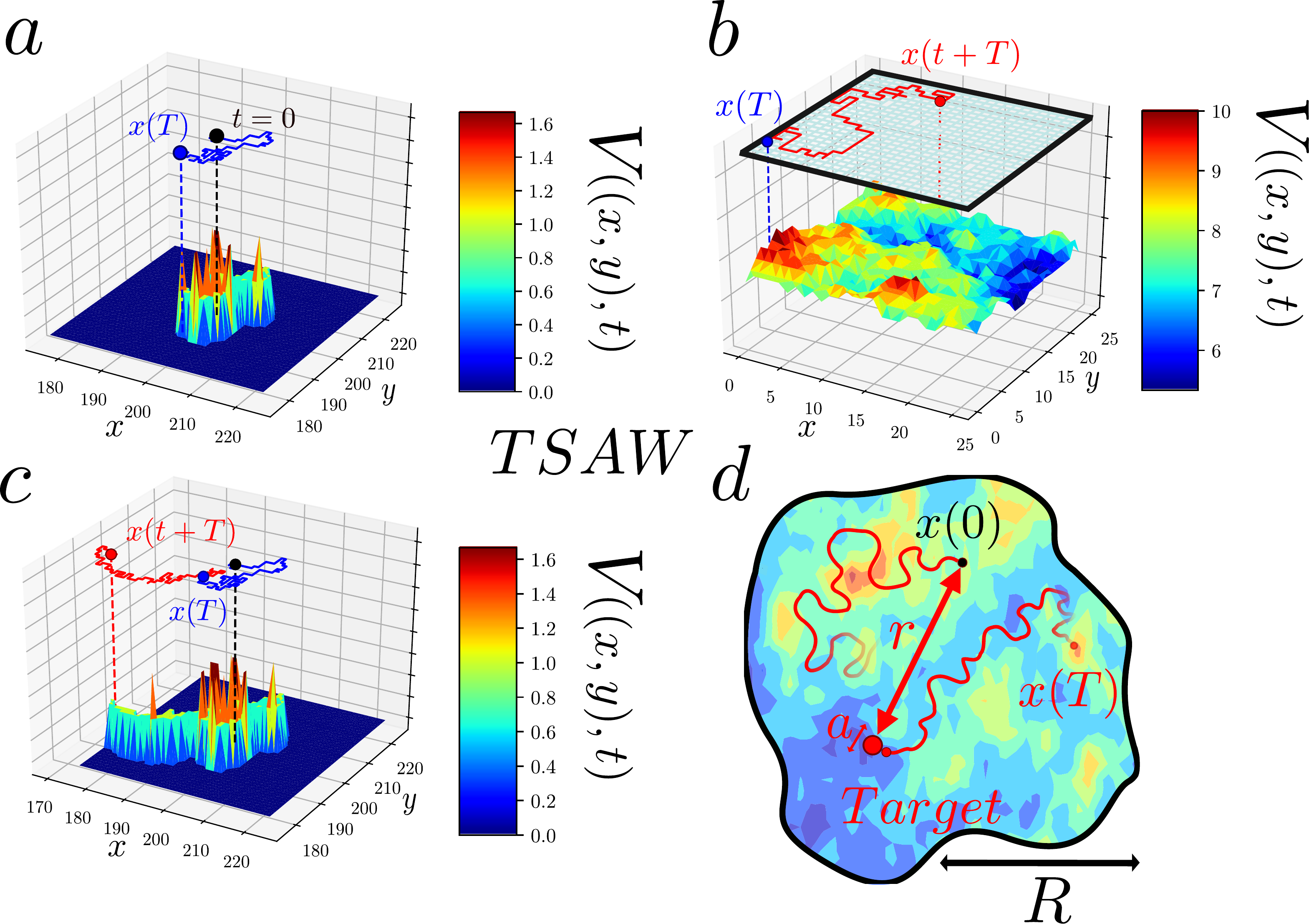}
    \caption{Aging and first-passage times for self-interacting random walks in infinite space and in confined domains. Examples of $2d$ self-interacting trajectories  (TSAW model, see text) \textbf{a.} The random path from $t=0$ to $T$  generates the local energy landscape $V((x,y),t)$ (proportional to the total number of visits to $(x,y)$ until $t$), plotted along the vertical axis $z$. This dynamics can lead to long range memory effects and aging at all time scales : the trajectory  after $T$ plotted in \textbf{c.}  explicitly depends on the full territory visited by the walker until $T$. In confined domains \textbf{b.}, the statistics of visits to a given site is radically modified by confinement : the dynamics of self-interacting random walks is thus geometry dependent. \textbf{d.} In this article, we aim at quantifying  space exploration and search kinetics of confined self-interacting random walks.}
    \label{fig1}
\end{figure}

More precisely, self-interacting random walks can be defined as nearest neighbor random walks on a $d$-dimensional lattice,  for which  the probability to jump to a neighboring site $i$ at time $t$ is proportional  to a weight function $w(n_i)$ that  depends on the number of previous visits  $n_i$ of the random walker to site $i$ up to time $t$ (see Fig.\ref{fig1}). %Note that alternatively the variable $n$ can be defined on bonds instead of sites ; this prescription will  be called bond convention   below, and will not be discussed in details.  Denoting by $x_t$ the position of the random walker at time $t$ and by $V(x_t)$ the set of neighboring sites,
% the transition probability defining the process can be written for $k\in V(x_t)$: 
%\begin{equation}\label{trans}
%p( x_{t+1} = k |  \{x_{t},...,x_{0}\}) = \frac{w(n_k)}{ \sum_{j\in V(x_t)} w(n_j) }.
%\end{equation}
 Writing $w(n)=e^{-V(n)}$, the process has the following clear interpretation : upon visiting site $i$, the random walker deposits a signal that in turn modifies the local energy landscape $V$ experienced by the walker. Of note, in contrast to autochemotactic or autophoretic systems \cite{Pohl:2014uh,Theurkauff:2012ui}, the deposited signal is assumed to be static and permanent, but not diffusive, which leads as we argue below to long lived memory effects.      To cover a broad spectrum of possible behaviours, we will consider both attractive ($V$ decreasing) and repulsive ($V$ increasing) self-interacting random walks, with effective potentials ranging from linear ($V(n)=\beta n$) to bounded ($V(n)=\beta H(n)$), where $H(n)$ denotes the Heaviside function \cite{davis_reinforced_1990,pemantle_survey_2007,toth_self-interacting_2001,stevens_aggregation_1997,foster_reinforced_2009,ordemann_structural_2001}.

Despite their relevance in various contexts, the properties of self-interacting random walks remain poorly understood, even if significant results have been obtained in the mathematical \cite{ pemantle_survey_2007,toth_self-interacting_2001, davis_reinforced_1990, horvath_diffusive_2012} and physical communities \cite{amit_asymptotic_1983,ottinger_generalised_1985,sapozhnikov_self-attracting_1994,foster_reinforced_2009, ordemann_structural_2001, grassberger_self-trapping_2017, freund_how_1993, peliti_random_1987,Kranz:2016vy,Kranz:2019uw,BarbierChebbah:2020aa,Campos:2021tm}. This stems from the strongly non Markovian nature of self-interacting random walks, whose dynamics depends on the set  of number of visits (or local times)  $\{n_i\}_{i\in\mathbb{Z}^d}$ at all sites $i$ of the lattice at time $t$, and therefore  on the full trajectory $ \{X(t')\}_{t'\le t}$ of the random walker up to time $t$. This dependence leads to memory effects at all time scales, which  can have important consequences depending on the potential $V(n)$ and space dimension $d$, such as  anomalous diffusion -- defined as the anomalous scaling of the mean squared displacement (MSD) :
\begin{equation}\label{msd}
 \langle X^2(t) \rangle  \underset{t \rightarrow \infty}{ \propto } t^{2/d_w}
\end{equation} 
with the walk dimension  $d_w\not=2$, or aging -- that can be defined  as the dependence of increments 
\begin{equation}\label{inc}
 \Delta^2(T,t)\equiv\langle (X(T+t)-X(T))^2 \rangle  \equiv2 D(T,t) t^{2/dw}
 \end{equation} 
 on the observation time $T$, where $D(T,t)$ is the effective time dependent diffusion coefficient. As we recapitulate below, the analytical determination of  $d_w$ remains a theoretical challenge ; so far it has been obtained analytically or numerically for different  examples of $V(n)$ and $d$, but even its numerical determination remains debated for attractive linear $V(n)$ for $d=2$. In turn, the  aging properties of  $\Delta^2(T,t)$ have not been studied until recently \cite{BarbierChebbah:2020aa} and will be analyzed in this paper.

A central question that  arises in random walk theory is the quantification of space exploration  by a  random walker \cite{Redner:2001a,bookSid2014,Benichou:2011fk,Viswanathan:2008}. Beyond the MSD and increments of the position, which provide a first quantification of  the dynamics of spreading in space, several observables have been proposed to quantify space exploration. Among those,   the first-passage time (FPT) and its distribution have proved to play a key role \cite{Redner:2001a,bookSid2014,Condamin:2007zl,BenichouO.:2010,Mattos:2012ys,Benichou:2014fk,Guerin:2016qf,Levernier:2018qf,Giuggioli:2020tf}. Indeed, beyond being a prominent technical tool of random walk theory that gives access to various  observables,  it quantifies the kinetics of general target search problems at all time scales, and as such  has a broad range of applications  from diffusion limited reactions to animal foraging  behaviour.

In infinite space, the first-passage statistics to a target follow two very distinct behaviours depending on the so-called type of the random walk \cite{Hughes:1995}. In the compact or recurrent case,   the  survival probability $S(t)$, \emph{i.e.} the probability that the target has not been found until  time $t$  typically vanishes at long time scales  as 
\begin{equation}\label{theta}
S(t) \underset{t \rightarrow \infty}{ \propto } t^{-\theta},
 \end{equation}
  where $\theta$ is  the persistence exponent,  which  has been the focus of numerous studies  \cite{Bray:2013}. In the non compact or transient case, the survival probability  admits a non zero large time limit, which defines  the hitting probability $\Pi$ \cite{Hughes:1995,Redner:2001a} according to $S\displaystyle \mathop{\to}_{t\to\infty} 1-\Pi$. In turn, the hitting probability is expected to decrease with the  distance $r$ from the starting position of the random walk and  the target  radius $a$ according to 
 \begin{equation}\label{psi} 
 \Pi  \underset{r \rightarrow \infty}{ \propto } (a/r)^{\psi}.
 \end{equation} 
  The   corresponding transience exponent  $\psi$ was  recently introduced in \citep{Levernier:2018qf,Levernier:2021aa} and parallels the persistence exponent of recurrent processes.  In spite of their pivotal role in quantifying first-passage properties of random walks, determining analytically the exponents $\theta, \psi$ for general non Markovian, aging processes remains a theoretical challenge  \cite{Bray:2013,Levernier:2019aa}. In particular, they remain unknown analytically for most examples of self-interacting processes, with the exception of \cite{BarbierChebbah:2020aa} ; they will be analyzed numerically and analytically in this paper.

In the case of geometrically confined spaces, which is relevant to most of practical situations where the space accessible to the random walker is ultimately bounded, space exploration  is known to be radically different. In particular, a target is eventually found with probability one for both compact and non compact processes, and the broad tails of the FPT distribution are generally suppressed \cite{BenichouO.:2010,Meyer:2011,Mattos:2012ys,Levernier:2018qf}. FPT statistics in confinement has been the subject of intense activity over the last decade, and general results have been obtained for general scale-invariant Markovian processes  \cite{BenichouO.:2010,Meyer:2011} or Gaussian non Markovian processes \cite{Guerin:2016qf}. Notably, more recently a universal scaling form of the FPT distribution was derived in the limit of large confining volume for a  class of scale-invariant non Markovian processes that display power law aging  \cite{Levernier:2018qf} ; importantly, the scaling of the FPT distribution in confinement was found in this case to be fully determined asymptotically by the exponents $d_w, \theta, \psi$ (in addition to the space dimension $d$), which are all defined {\it in infinite space} and independent of the geometric confinement.

Quantifying space exploration of self-interacting random walks in confined geometry brings in this context a new conceptual challenge. Indeed, qualitatively it is expected that geometric confinement will  modify  the statistics   of the numbers  of visits  $\{n_i\}$ at the  sites $i$ of the confined domain over time and thus the effective potential $V(n_i)$ experienced by the walker, thereby impacting the very dynamics of the process, as compared to that in infinite space. This can be illustrated by the simple example of the normal random walk for $d=3$ : in infinite space, the mean number of visits $\langle n_i\rangle$ to any site $i$ converges to a finite value for $t\to\infty$, whereas it diverges  as $\langle n_i\rangle  \propto  t$ in a confined domain. In the case of self-interacting random walks, the kinetics of space exploration  thus directly  feeds-back to the dynamics of the process {\it in a geometry dependent manner}, which suggests that key intrinsic features of the dynamics, such as increments (quantified by $D(T,t)$ and $d_w$), persistence and transience exponents $d_w,d_f$ could in fact be different in confined and infinite geometries. This in particular makes earlier approaches to determine FPT statistics inapplicable, and calls for new theoretical tools to quantify the space exploration  of confined self-interacting random walks ; this is at the core of this paper.

Our findings can be summarized as follows. We show that universal scaling forms of the FPT distributions of general self-interacting random walks in confinement  can be derived in the large volume limit, by generalizing the approach introduced in \citep{Levernier:2018qf}.  Because of the intrinsic aging properties of self-interacting random walks, different cases emerge
depending on the preparation protocol. For "fresh" initial conditions, for which the random walker starts the search for a target  in a domain that has never been explored,  we find that the  exponents $d_w, \theta, \psi$ that determine the FPT distribution are generally identical to those 
 defined in infinite space : in other words, the FPT distribution in confinement can be asymptotically predicted from the knowledge of he process in infinite space only. This is quite  remarkable because, as we show,  geometric confinement ultimately deeply modifies the dynamics of the process and can even change  the corresponding   exponents $d_{w,c}, \theta_c, \psi_c$  defined in confinement.
In contrast, for aged initial conditions, for which the random walker has been extensively wandering   in the domain before the search starts, the  exponents that determine the FPT distribution are those defined in confinement $d_{w,c}, \theta_c, \psi_c$ , and can thus be different from the classical infinite space exponents $d_w, \theta, \psi$. In that case, the process in confinement must therefore be characterized to determine the FPT distribution. In all cases, scaling functions  are not universal and are  process dependent. This analysis is made possible by a systematic quantitative characterization of the aging properties (quantified by $D(T,t)$ and $d_w$) and  exponents   $\theta$ or $\psi$ of self-interacting random walks in both confined and infinite geometries, which highlights  the impact of geometric confinement on their dynamics.  Finally, this paper thus proposes a unified, quantitative analysis   of aging, exploration and FPT statistics of self-interacting random walks in confined and infinite geometries.

The paper is organized as follows. First, we briefly define the main classes of attractive and repelling self-interacting walks and recall their walk dimension $d_w$ when it is known. In particular we provide a general criterion for attractive walks  that leads to bounded exploration ($d_w=\infty$), or to the regime of strong attraction ($\theta=\infty$, to be defined below) ; in this case the FPT problem in confinement is trivially equivalent  in the large volume limit to the problem in infinite space. Second, we characterize quantitatively the increments and their aging behavior, as well as the persistence and transience exponents $\theta,\psi$ in both confined and infinite geometries. Third, based on this analysis, we derive the asymptotic FPT distribution  in confinement for both compact and non compact self-interacting random walks for fresh initial conditions.  Last, we discuss the impact of aging on FPT statistics by analyzing the case of aged initial conditions, which allows us to assess the impact of memory effects on target search kinetics of self-interacting random walks.

\section{Definitions and main classes of self-interacting random walks}

As stated above, self-interacting random walks can be defined as nearest neighbor random walks on a $d$-dimensional lattice,  for which  the probability to jump to a neighboring site $i$ at time $t$ is proportional  to a weight function $w(n_i)$ that  depends on the number of previous visits  $n_i$ of the random walker to site $i$ up to time $t$. %Note that alternatively the variable $n$ can be defined on bonds instead of sites ; this prescription will  be called bond convention   below, and will not be discussed in details.  Denoting by $x_t$ the position of the random walker at time $t$ and by $V(x_t)$ the set of neighboring sites,
% the transition probability defining the process can be written for $k\in V(x_t)$: 
%\begin{equation}\label{trans}
%p( x_{t+1} = k |  \{x_{t},...,x_{0}\}) = \frac{w(n_k)}{ \sum_{j\in V(x_t)} w(n_j) }.
%\end{equation}
%Note that the set of local times  (or number of visits)  $\{n_i\}_{i\in\mathbb{Z}^d}$ at time $t$ depends on the full trajectory $ \{x_{t},...,x_{0}\}$ up to time t. The process is therefore strongly non Makovian and has long range memory. 
Denoting $w(n)=e^{-V(n)}$, the process has the following clear interpretation : upon visiting site $i$, the random walker deposits a signal that in turn modifies the local energy landscape $V$ experienced by the walker.   
Different classes of random walks are obtained depending on the choice of weight function $w(n)$ ; we remind below the main known results concerning the MSD of these processes.

  %To cover a broad spectrum of possible behaviours, we will consider the following representative examples of weight functions  $[w(n) = e^{-\beta n}, \beta < 0]$, subexponential $[ w(n) = e^{-\beta n^k}, \beta < 0, 0<k<1]$, polynomial $[ w(n) =n^\beta, \beta >  0]$ or asymptotically constant  $[ w(n) \sim 1$ for $n\to\infty]$ weight functions [\cite{davis_reinforced_1990,pemantle_survey_2007,durrett_asymptotic_1992,toth_self-interacting_2001,stevens_aggregation_1997,foster_reinforced_2009,ordemann_structural_2001}]

\subsection{The True Self Avoiding Walk  (TSAW) : $w(n)\propto e^{-\beta n}$}
In this  model \cite{amit_asymptotic_1983,Pietronero:1983vo,obukhov_renormalisation_1983,toth_self-interacting_2001} the effective potential $V(n)$ depends linearly on the local time  $n$. For $\beta<0$ the interaction is attractive, and leads (almost surely), as we show below, to the  complete trapping of the random walker on a finite set of sites for all $d$, and thus formally to $d_w=\infty$:
\begin{equation}
\langle X(t)^2\rangle \underset{t \rightarrow \infty}{ \to } C,
\end{equation}
where $C$ is a $d$ dependent constant.
 For $\beta>0$ the interaction is repulsive and the random walker qualitatively  avoids its own path. It has been shown that this leads to the following scaling of the  MSD for $t\to \infty$ \cite{amit_asymptotic_1983,Pietronero:1983vo,obukhov_renormalisation_1983,grassberger_self-trapping_2017}:
\begin{itemize}
\item $d=1: \: \:$ $\langle X(t)^2\rangle \: \propto \: t^\frac{4}{3},  \: \: \:d_w = \frac{3}{2}$
\item $d=2 :\: \:$ $\langle X(t)^2\rangle \:  \propto \: t \: \ln(t)^\alpha,  \: \: \:d_w = 2,  \:  \alpha \approx 0.5 $
\item $d=3 : \: \:$ $\langle X(t)^2\rangle \:  \propto \: t,  \: \: \: d_w = 2$
\end{itemize}
Of note, the scaling of the MSD is thus anomalous (superdiffusive) for $d\le 2$ because of self-repulsion, while it is diffusive for $d>2$.

%The TSAW is then compact in dimension 1 meaning that a target in infinite media is found with a probability 1, marginally compact in dimension 2 transient in dimension 3.
%For $\beta \rightarrow \infty$ (zero temperature) the walks tends to a process that systemitaically jump on the less visited site ( chosen uniformly along the less visited). This limit case is share with the SESRW. \medbreak

\subsection{The Sub-Exponential Self Repelling Walk (SESRW) : $w(n)\propto e^{-\beta n^k}$}
This model \cite{ottinger_generalised_1985,toth_self-interacting_2001,toth_true_1995} extends the TSAW to effective potentials $V(n)$ that depend sublinearly on the local time $n$: $V(n)=\beta n^k$ with $0<k<1$. Similarly to the TSAW, in the attractive case ($\beta<0$), the random walker  is (almost surely) trapped for all $d$, and thus $d_w=\infty$.  For $\beta>0$, the effect of self avoidance is clearly weaker than for the the TSAW ; it has however been shown to still lead to superdiffusion for $d=1$ \cite{ottinger_generalised_1985,toth_true_1995}. This can be summarized as follows : 
\begin{itemize}
\item $d=1: \: \: \langle X(t)^2\rangle\:\propto \: t^\frac{2(1+k)}{(2+k)},  \: \: \: d_w = \frac{2 + k}{1 + k}$
\item $d=2: \: \:$ $\langle X(t)^2\rangle \: \propto \: t \: \ln(t)^{\alpha_k},  \: \: \:d_w = 2,  \: \:$ $ \: \: \alpha_k\ge 0 $
\item $d=3: \: \:$ $\langle X(t)^2\rangle \: \propto \: t,  \: \: \: d_w = 2$
\end{itemize}

\subsection{The self attractive random walk (SATW) : $w(n)\propto e^{-\beta H(n)}$}
In this model \cite{sapozhnikov_self-attracting_1994, prasad_diffusive_1996,  pemantle_survey_2007,foster_reinforced_2009, ordemann_structural_2001, davis_reinforced_1990, agliari_true_2012}, the effect of self interaction is assumed to saturate with the number of visits, so that the effective potential $V(n)$ is bounded for $n\to\infty$. For the sake of simplicity, it is assumed in the SATW model  that $V(n)=\beta H(n)$, with $H(0)=0$ and $H(n\ge 1)=1$. Note that the SATW can thus be seen as the $k \to 0$ limit of the SESRW defined above.
For $\beta>0$, self-avoidance is insufficient to modify the scaling of the MSD, which remain diffusive for all $d$ : \newline
\begin{equation}
\langle X(t)^2\rangle \:\propto \: t,  \: \: \: d_w = 2.  
\end{equation} 
In the attractive case $\beta  < 0$, the random walker is never trapped.  For $d=1$ the  MSD satisfies \cite{sapozhnikov_self-attracting_1994, prasad_diffusive_1996} : 
\begin{equation}
\langle X(t)^2\rangle \:\propto \: t,  \: \: \: d_w = 2 , 
\end{equation} 
while for $d=3$ different behaviors emerge depending on the value of the parameter $\beta$
\begin{itemize}
\item  $ |\beta| < |\beta_c|:  \: \:$ $\langle X(t)^2\rangle \: \propto \: t,  \: \: \: d_w = 2$
\item  $ |\beta| > |\beta_c|: \: \:$ $\langle X(t)^2\rangle \: \propto \: t^{1/2},  \: \: \: d_w = 4$.
\end{itemize}
For $d=2$, the scaling of the MSD  is still debated \cite{ordemann_structural_2001,foster_reinforced_2009}; while the existence of a subdiffusive regime with $d_w = 3$ is consistently observed numerically,  the existence of a transition for a   critical value $\beta'_c\not=0$ to a diffusive regime with  $ d_w = 2$  for $ |\beta| < |\beta'_c|$ has been proposed, but was later questioned in \cite{foster_reinforced_2009}. 

\begin{figure}[h!]
    %\centering
    \includegraphics[scale=0.3]{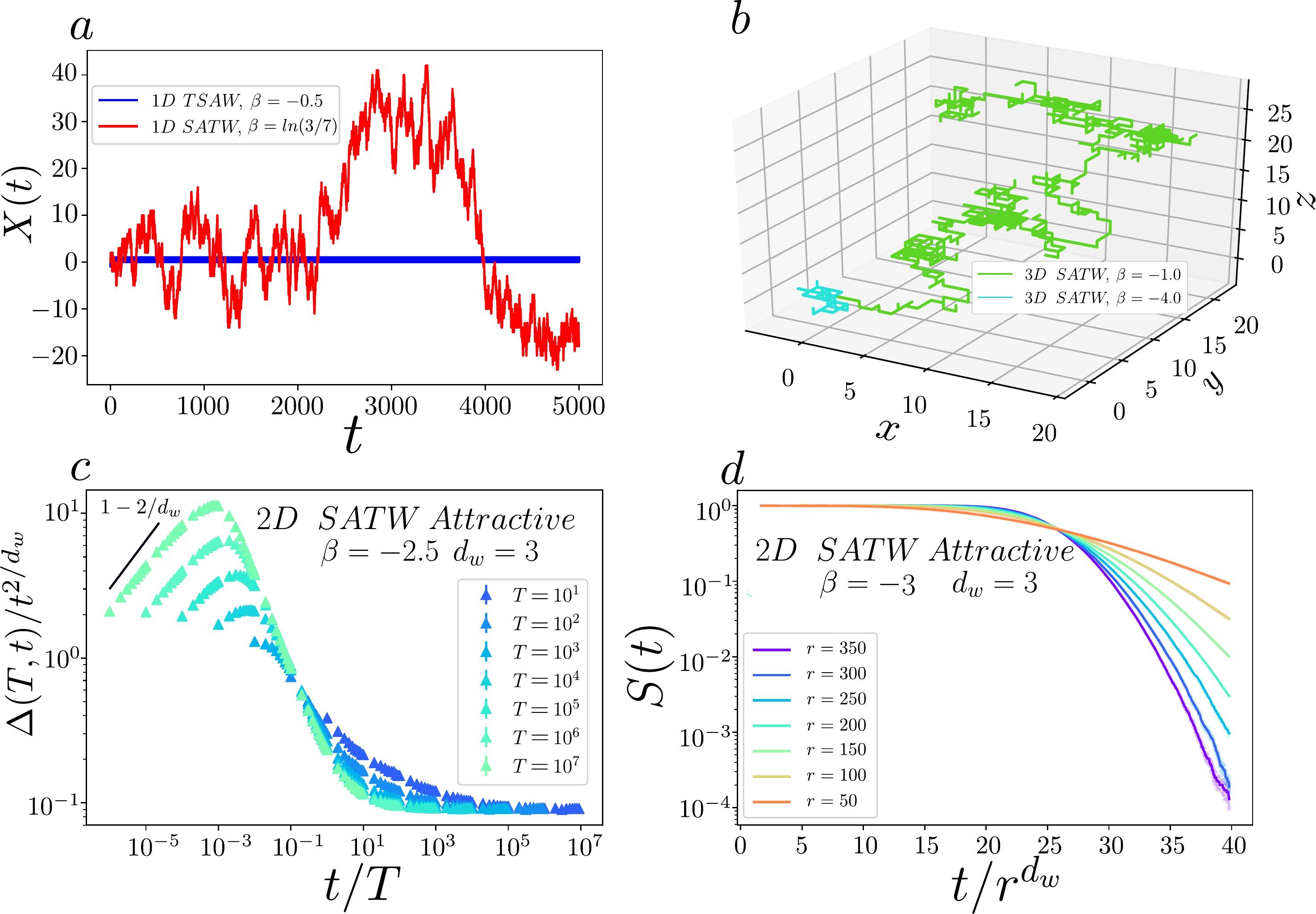}
    \caption{ General properties of attractive self-interacting random walks. \textbf{a.} Example of trapped trajectory performed by a $1d$ attractive TSAW random walker (blue), compared  to a diffusive trajectory of a $1d$  attractive SATW random walker (red). \textbf{b.} For $d>1$, the SATW is subdiffusive for  $|\beta| > |\beta_c|$ (blue sample trajectory), and diffusive for  $|\beta| < |\beta_c|$ (green sample trajectory). Here $d=3$ and thus $|\beta_c|\not=0$ \cite{foster_reinforced_2009}. \textbf{c.} Aging of the increments for the subdiffusive SATW ($d_w=3$ for  $d=2$) normalized by the expected subdiffusive scaling at long times. Each curve corresponds to a fixed value of $T$. Note that the increments are diffusive for $t \ll T $.  \textbf{d.} In the subdiffusive regime,  the SATW performs an extremely  compact exploration of space : the survival probability $S(t)$ decays faster than any powerlaw ($\theta=\infty$).
}
    \label{fig2}
\end{figure}

\section{attractive self-interacting random walks: trapping and subdiffusion }

Qualitatively,  attractive self-interacting random walks are attracted by their own path. Strikingly, this can lead to the full trapping of the walker within a finite set of sites in the $t\to\infty$ limit, and therefore to a bounded MSD.  This effect was demonstrated mathematically   \cite{davis_reinforced_1990,volkov_phase_2006} for 1--dimensional attractive self-interacting random walks and later generalized to arbitrary $d$ \cite{stevens_aggregation_1997,cotar_edge-_2017,sellke_reinforced_2008,basdevant_localization_2014} : %with the bond convention.
more precisely, these results state that if  $ \sum_{n=1}^{\infty} w(n)^{-1}= \infty  $ the random walker is  free and will visit infinitely many sites of the lattice (note that the limit case $w(n)\propto 1/n$ must be discussed independently). Conversely,  for $ \sum_{n=1}^{\infty} w(n)^{-1} < \infty $,  the random walker visits only a finite set of sites  and the MSD is bounded (see Supplementary Information SI).  %We argue in appendix that this criterion also holds for   $d$--dimensional self-interacting random walks. 
This yields immediately that attractive  TSAW and SESRW  lead to the full trapping of the random walker and to a bounded MSD for all $d$ (see Fig.\ref{fig2} ).  Among  the classes of attractive self-interacting random walks introduced above, the only case that leads to a non trivial exploration of space is thus the SATW, for which the MSD diverges for $t\to\infty$ (see Fig.\ref{fig2}). 
Despite the diverging MSD, the effect of attractive self interactions can still have important consequences on the dynamics of space exploration; in particular, for $d=2$ and $d=3$ (for $|\beta|>|\beta_c|$) the process is subdiffusive \cite{foster_reinforced_2009} and we find that  the survival probability  $S(t)$ in infinite space   decays faster than any power-law, so that $\theta=\infty$  (see Fig.\ref{fig2} and SI).
In this case,  determining  the FPT distribution starting at a distance $r$ from the target  $F(t,r,R)$ in confined domains of volume $V\propto R^d$ is straightforward in the large volume limit because all moments of $F(t,r,R)$ have a finite limit, so that: 
\begin{equation}
F(t,r,R)\underset{t \rightarrow \infty}{ \sim }-\frac{dS}{dt}.
\end{equation}
$F(t,r,R)$ is thus asymptotically independent of $R$. Defining the rescaled variable $\eta=t/r^{d_w}$,   a scaling argument finally indicates that its asymptotic distribution can be simply written:
\begin{equation}\label{Fbar0}
 {\bar F}(\eta,r,R)=  h(\eta)
\end{equation} 
where $h$ is an undetermined scaling function.  In the rest of this paper, we thus focus on diffusive attractive and all repulsive self-interacting random walks, for which determining the FPT distribution $F(t,r,R)$ in confined domains is non trivial.

\section{Impact of confinement on  increments, $\theta,\psi$ }

In this section, we characterize quantitatively the exploration properties of diffusive attractive self-interacting random walks and repulsive self-interacting random walks. We focus on the following observables : increments, and survival probability characterized by $\theta$ (for compact processes) and $\psi$ (for non compact processes) in both infinite and confined geometries. We show numerically and provide heuristic arguments to justify that geometric confinement can deeply and non locally modify the dynamics of the process, beyond imposing locally reflecting boundary conditions. As can be expected, it is useful to analyse separately compact and non compact  processes. While this property is known to impact  many properties of random walks, it is expected to play a prominent role in the case of self-interacting random walks, whose dynamics is controlled by the number of visits $n$ at each site.

\subsection{Compact (recurrent) processes}
The compact case is exemplified by the $1d$ (repulsive) TSAW, the $1d$ (repulsive) SERW and the $1d$ (attractive or repulsive) SATW. In the compact case, the mean number of visits $\langle n_i\rangle$ to a given site diverges with time $T$ by definition even in infinite space. The local energy landscape $V(n_i)$ experienced by the random walker  therefore  depends on the observation time $T$ at all time scales. We argue below that this leads to  aging of the increments in infinite space at all time scales, ie a dependence on $T$ of the effective diffusion coefficient $D(T,t)$ defined in \eqref{inc} for all $T$. In a confined domain, the dynamics of the random walk starting typically from the bulk is not modified by confinement up to an observation time $T\sim R^{d_w}$, where $R$ is the  typical linear size of the domain; in this regime we therefore expect the increments to be identical in both confined and infinite geometries (note that for the same reason, in confined domains the analysis of increments is restricted to $t\ll R^{d_w}$).  For $T\gtrsim R^{d_w}$, confinement does modify the statistics of visits to a given site; however  the number of visits to a given site still diverges with time $T$, even if the explicit dependence on $T$ is different in confinement and in  infinite space. Aging of the increments is thus expected in confinement as well.

To make this analysis quantitative, it is  useful to write $V(n_i)$ as a Taylor series:
\begin{equation}\label{taylor}
 V(n_i)=V({\bar n})+\sum_{p\ge 1}V^{(p)} ({\bar n})\frac{(\delta n_i)^p}{p!}, 
 \end{equation}
 where ${\bar n}$ denotes the number of visites to a given site  averaged over a spatial scale $l\ll T^{1/d_w}$ and $n_i={\bar n}+\delta n_i$.
The very definition of the dynamics of self-interacting random walks shows that  it depends  only on the spatial fluctuations  of $V$ ; increments are thus independent of the site independent contribution   $V({\bar n})$ for $t\lesssim l^{d_w}\ll T$. In the case of the SATW and the SESRW, one has $V^{(p)} ({\bar n}) \propto {\bar n}^{k-p}$ by definition (we remind that for the SESRW $V\propto n^k$, where $k\to 0$ yields the SATW) ; in addition, a mean field argument (see SI) yields the scaling $\delta n_i\propto {\bar n}^{(1-k)/2}$.
This shows that  in the regime $1\ll t\ll T$  all site dependent terms $V^{(p)} ({\bar n})(\delta n_i)^p/p!$ for $p\ge 1$ appearing in \eqref{taylor} vanish in the limit $T\to \infty$ in both confined and infinite geometries because ${\bar n} \to \infty$. Self-interactions are thus eventually negligible in this limit: the SATW and the SESRW  are   equivalent   to a simple random walk and one has identically  $\Delta^2(T,t)\sim t$ in both confined and infinite geometries. In the case of the TSAW, one has $V' ({\bar n})=\beta$ and $V^{(p)} ({\bar n})=0$ for $p>1$, independently of ${\bar n}$.   Using in addition the fact that the spatial fluctuations $\delta n_i$ reach a steady state in the limit ${\bar n}\to\infty$ (see \cite{freund_how_1993} and SI), this shows that in the regime $1\ll t\ll T$ the dynamics of increments is identical in both confined and infinite cases ; it can be shown to satisfy $\Delta^2(T,t)\propto t^{2/d_w}$. Last, in the regime $1\ll T\ll t $, one recovers the scaling of the MSD  in all cases :  $\Delta^2(T,t)\propto t^{2/d_w}$.

\begin{figure*}
    \centering
    \includegraphics[scale=0.84]{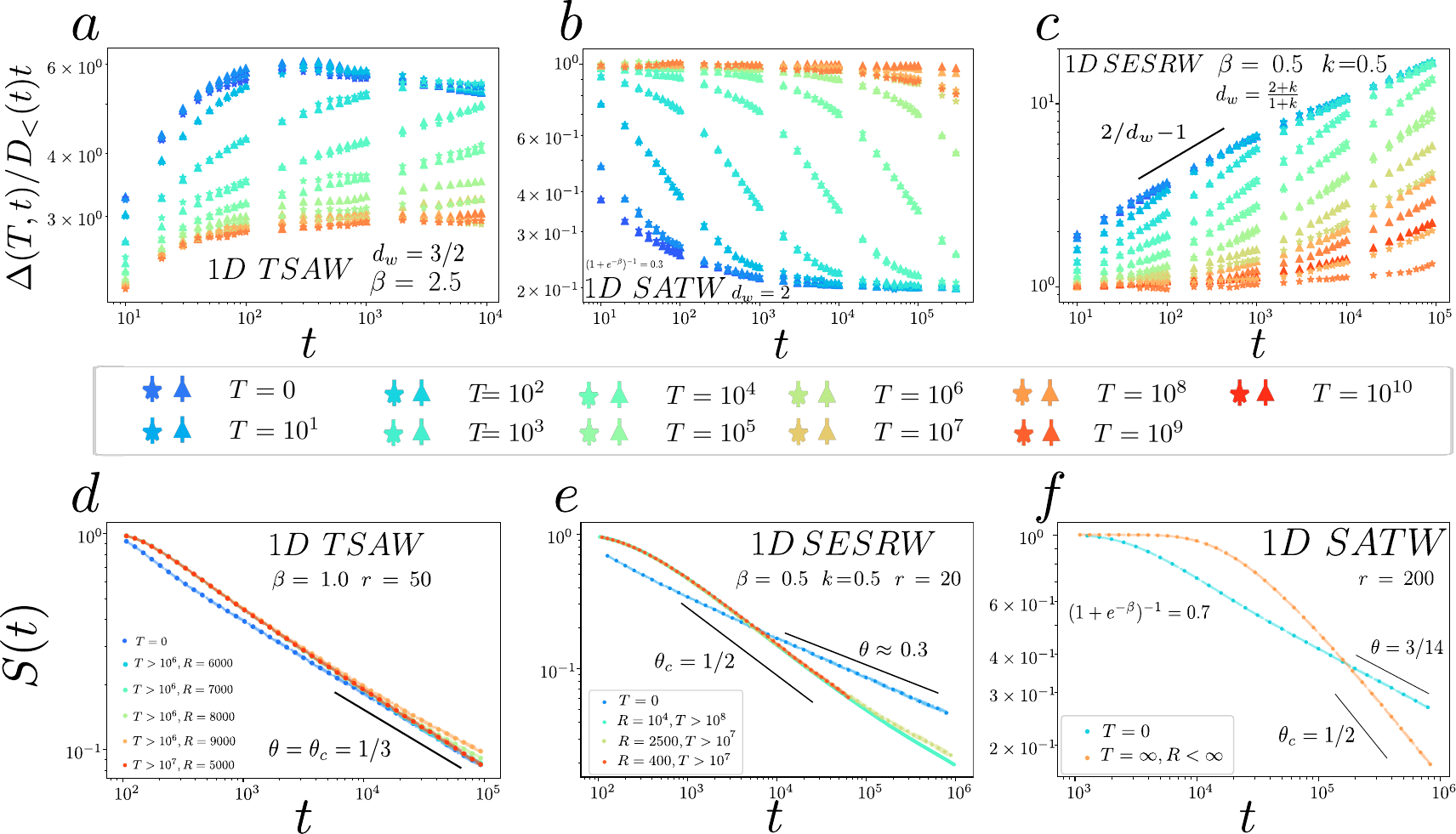}
    \caption{aging and first-passage properties for compact self-interacting random walks in infinite space and in confined geometries. aging of the increments for the $1d$ TSAW (\textbf{a.}), the $1d$ SATW (\textbf{b.}) and the $1d$ SESRW (\textbf{c.}) in both infinite space and confined domains  (increments are normalized by the expected scaling for $T \gg t$). The dynamic is found to be identical in infinite space (triangle) and in confined domains (stars) in both regimes $t,T\ll R^{d_w}$ and $T \gtrsim R^{d_w}, t\ll R^{d_w}$. The TSAW is superdiffusive at all times scales and displays aging ; the SATW is diffusive at all times scales and displays aging. In contrast, the SESRW is diffusive for $T \gg t$, but superdiffusive for $t \gg T$  with $d_w = (2+k)/(k+1)$. Persistence exponents in infinite space ($\theta$) and in confined domains  ($\theta_c$) for the $1d$ TSAW (\textbf{d.}),  the $1d$ SESRW (\textbf{e.}) and for the $1d$ SATW (\textbf{f.}). The persistence exponent is modified by confinement for the SATW and the SESRW models, but is found numerically to be unchanged for the TSAW.}
    \label{fig3}
\end{figure*}

These results can be recapitulated for all examples by the following scaling forms, which are identical in confined and infinite geometries :
\begin{eqnarray}
1\ll t \ll T \ : & \ \Delta^2(T,t)\sim 2 D_<(t) t^{2/d_w}\nonumber\\
1\ll T\ll t \ : & \Delta^2(T,t)\sim 2 D_> t^{2/d_w},
 \end{eqnarray}
where the constant $D_>$ and function $D_<(t)$ are process dependent. Numerical simulations confirm this analysis in all examples of compact self-interacting random walks (see Fig.\ref{fig3}) : increments display aging  (as seen by a dependence of $\Delta^2$ on the observation time $T$), and their dynamics is  found to be the same  in infinite space and in confined domains in both regimes $t,T\ll R^{d_w}$ and $T \gtrsim R^{d_w}, t\ll R^{d_w}$.  %At this time scale $t\ll R^{d_w}$, aging of the increments occurs over a finite range of observation time $T< T_a$,  while increments   become independent of $T$ for $T>T_a$. 
%XXXXXXX  EXPLAIN ?? TRIVIAL for SATW, ARGUMENT STATIONARY VARIANCE FOR TSAW ?????   XXXXXXX
%XXXXXXXXMention that for infinite space , aging at all time scales XXXXXXXXX

In contrast to the dynamics of  increments,  we now argue that the persistence exponent $\theta$  can be modified   by confinement. Following \cite{Krug:1997}, we introduce here the persistence exponent $\theta_c$ in confinement that can be defined by 
\begin{equation} \label{eqsurvival}
S(t|T)\propto t^{-\theta_c}
\end{equation}
for $T\gg R^{d_w}$  and $1\ll t \ll R^{d_w}$, where $S(t|T)$ denotes the (survival) probability that the  random walker has not reached the target between $T$ and $T+t$.  It is known  that $\theta $ depends on the dynamics of increments at all time scales \cite{Bray:2013}, and not only on their long time asymptotics. The exponents $\theta$ and $\theta_c$ can thus be different, as was earlier found in \cite{Krug:1997} for models of fluctuating interfaces,  because $\theta$ involves the dynamics of increments at all time scales $T,t$, while the definition of $\theta_c$ only involves the time scales $t \ll R^{d_w}$ and $T\gg R^{d_w}$. This is straightforwardly confirmed in the case of the $1d$ SATW. It is clear that for $T\gg R^{d_w}$, the confined SATW is equivalent to a simple random walk (in this regime all sites have been visited and $V(n_i)=\beta$ for all sites), so that  $\theta_c=1/2$; in contrast, it was shown recently that in infinite space one has   $\theta=e^{-\beta}/2$ \cite{BarbierChebbah:2020aa}.  In the case of   the SERW, the above analysis shows that in the regime $T\gg 1$, the process is also equivalent to a simple random walk, so that $\theta_c=1/2$ ; in contrast, we find numerically $\theta\not=\theta_c$. Note however that    it is found numerically that $\theta_c\approx\theta\approx 1-1/d_w$ for  the TSAW for all $\beta>0$.

\subsection{Non compact (transient) processes}
\begin{figure*}
    \centering
    \includegraphics[scale=0.94]{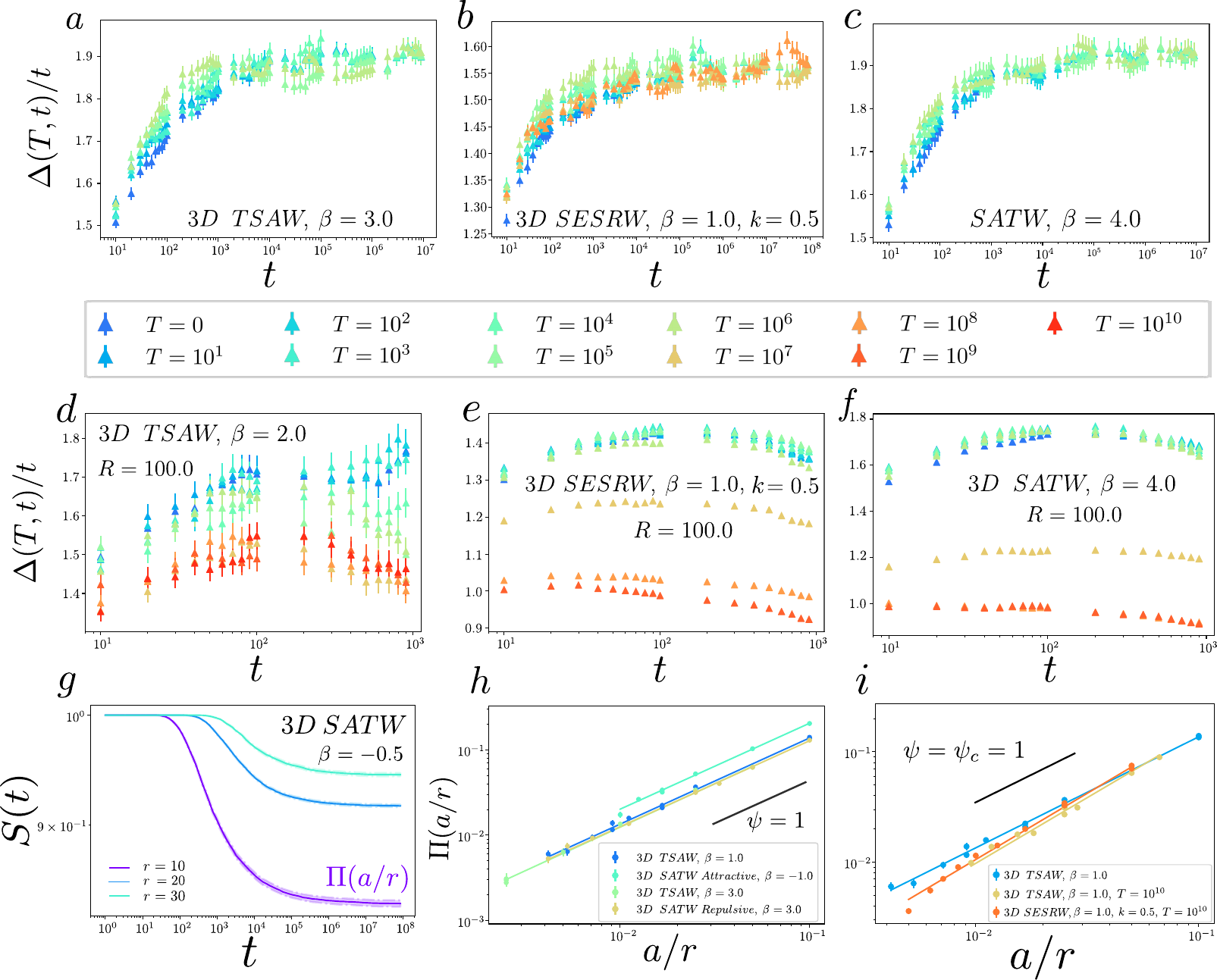}
    \caption{Aging and first-passage properties for non-compact self-interacting random walks in infinite space and in confined geometries. Aging of the increments for the $3d$ TSAW (\textbf{a.}), the $3d$ SESRW (\textbf{b.}) and the $3d$ SATW (\textbf{c.}) in infinite space (increments are normalized by the expected diffusive scaling at long times). Of note, the increments are stationary at  timescales $T\gg 1$. In contrast,  in confined geometries, aging occurs at longer time scales $\gtrsim R^d$,  for the TSAW (\textbf{d.}), the SESRW (\textbf{e.}) and the SATW (\textbf{f.}) (increments are normalized by the expected diffusive scaling at long times). \textbf{g.} For non-compact random walks  (here the $3d$  diffusive SATW), the survival probability tends for $t\to\infty$ to a non-zero value $1-\Pi$, which defines the hitting probability that depends on the initial distance to the target $r$ and the target radius $a$. Hitting probability and  transience exponent  in infinite space (\textbf{h.}) and in confined domains (\textbf{i.}). Numerical simulations (symbols) and power law fits (plain lines). Our numerical results indicate  $\psi =\psi_c=1$ for the TSAW, the SESRW and the SATW, in agreement with the asymptotic diffusive behaviour of non compact self-interacting random walks.  }
    \label{fig4}
\end{figure*}

The non compact case is exemplified by the $3d$ (repulsive) TSAW, the $3d$ (repulsive) SERW and the $3d$ (diffusive attractive or repulsive) SATW. In the non compact case, in infinite space, a random walker visits only a fraction of sites, and ultimately only makes on average a finite number of visits  to a given site $i$. The local energy landscape $V(n_i)$  therefore  reaches a stationary state  at large observation time $T$, so that aging, if any,  is expected to be transient : $D(T,t)$ is asymptotically independent of $T$ for $T\gg 1$. This is indeed observed numerically in all examples of non compact self-interacting random walks:  increments  display weak aging at short time scales $t$, and cross-over to diffusive increments with numerically close diffusion coefficients at larger $t$ for all observation times $T$ (see Fig.\ref{fig4}). The effect of sel-interaction is thus moderate for non compact self-interacting random walks, which are all eventually diffusive. This can be heuristically  justified as follows:  at time $t$, the typical volume covered scales as $t^{d/d_w}$, while the number of visited sites scales as $t$, so that the local fraction of sites where   the local energy landscape is non zero eventually vanishes for $t\to\infty$ as $t^{1-d/d_w}$. Self interactions are thus negligible in the large time limit for non compact processes, which are diffusive in this limit (note however that the diffusion coefficient is non trivial and depends on the small $t$ dynamics).

The case of confined geometries is radically different for non compact processes, because confinement   leads to a divergence of the number of visits to a given site, and  has thus important consequences at time scales $T\gtrsim R^{d_w}$. In this regime, the above reasoning developed after \eqref{taylor} for compact processes in fact  applies also to confined non compact processes, because the locally averaged  number of visits ${\bar n}$ diverges in both cases. In particular, this yields similarly  that  in the regime $1\ll t\ll T$  both the non compact SATW and the non compact SESRW are  equivalent to a simple random walk, so that    $\Delta^2(T,t)\sim t$.
In the case of the confined non compact TSAW, one finds numerically (see also \cite{amit_asymptotic_1983,horvath_diffusive_2012} and SI for an heuristic argument) that the spatial fluctuations $\delta n_i$ reach a steady state in the limit $T\to \infty$. This, together with \eqref{taylor},  allows us to conclude that  in this limit increments are similar (scaling wise) to the infinite space case and thus diffusive and independent of $T$ (see SI). %Finally, for all confined non compact self-interacting random walks,  
%for $t\gg 1$ one can write $D(T,t)\simeq D_<$ for $T\ll R^{d_w}$ and $D(T,t)\simeq D_>$ for $T\gg R^{d_w}$, where $D_<,D_>$ are process dependent constants.

Finally, for all confined non compact self-interacting random walks, these results can be recapitulated as follows for $R\gg 1$:
\begin{eqnarray}\label{incnc}
T\lesssim R^{d_w}: & \Delta^2(T,t)\underset{T\gg 1}{\sim} 2 D_<(t)  t,\ D_<(t)\underset{t\gg 1}{\to} D_< \nonumber\\
T \gtrsim R^{d_w}: & \ \Delta^2(T,t)\underset{t\gg 1}{\sim} 2 D_>(T) t,\  D_>(T)\underset{T\gg 1 }{\to} D_>
\end{eqnarray}
where  $D_<,D_>$ are constants. The first regime $T\lesssim R^{d_w}$ is the same in confined and infinite geometries, while the second regime $T \gtrsim R^{d_w}$ is controlled by geometric confinement.  Numerical simulations confirm this analysis in all examples of non compact self-interacting random walks (see Fig.\ref{fig4}) : increments, even if  always asymptotically  diffusive, are found numerically in all examples to be quantitatively different for confined and non confined non compact self-interacting random walks.

Last, for the sake of completeness, we note that similarly to the persistence exponent in the compact case (see \eqref{eqsurvival}), the transience exponent $\psi_c$ can be defined in confinement according to :
\begin{equation}
  S(t|T) \underset{a \rightarrow 0}{\propto} \left(\frac{a}{r}\right)^{\psi_c} 
\end{equation}
for $T\gg R^{d_w}$ and $a^{d_w}\ll t \ll R^{d_w}$. While, in principle $\psi_c$ can be different from its infinite space counterpart $\psi$, our above analysis showed that  all examples of non compact self-interacting processes that we analysed are diffusive and independent of $T$ for $t\gg 1$ in the limit $T\to \infty$ in both confined and non confined cases; this suggests that $\psi=\psi_c=1$, which is consistent with our numerical simulations  (see Fig.\ref{fig4}).

To summarize this section, we have  showed quantitatively that geometric confinement can deeply and non locally modify the dynamics of self-interacting random walks, beyond imposing locally reflecting boundary conditions. In the compact case, increments remain unchanged (in  the regime $t\ll R^{d_w}$)  in confined and unconfined geometries, but the persistence exponent can be modified. In the non compact case, increments remain asymptotically diffusive in both cases, but their dynamics is   quantitatively modified by geometric confinement ; in turn, it is found that the transience exponent is unchanged.

%show increments that can be different confined vs non confined  [compare to FBM where confinement does not change anything  / identify time scale when it starts being different]

%Figure : show critical exponents that can be different confined vs non confined  [at least theta]

\section{FPT distribution in confined domains}
The above analysis of increments and exponents $\theta$ and $\psi$ shows that these observables can be impacted by confinement. Turning to the analysis of FPT properties of self-interacting random walks in confinement, one therefore needs to develop a new methodology. Indeed, so far, available methods to determine  FPT statistics in confinement \cite{Levernier:2018qf} rely implicitly on the hypothesis that increments and exponents $\theta$ and $\psi$, which are the key quantities defining the universality classes of FPT statistics in confinement, are not modified by confinement. Below, we extend the method developed originally in \cite{Levernier:2018qf} to the case of self-interacting random walks by taking explicitly into account the impact of confinement on the dynamics. In this section, we consider the case of "fresh" initial conditions : at $t=0$ the random walker, confined in a domain of volume $V=R^d$ with reflecting walls, starts at a distance $r$ from the target of radius $a$, and the number of visits to all sites $i$ of the domain is set to $n_i=0$. As stated in introduction, we focus on diffusive attractive and repulsive self-interacting random-walks,  and consider separately the cases of compact processes (for which the survival probability $S(t)$ has a power-law decay in infinite space) and non compact processes ; the case of marginal exploration ($2d$ processes with $d_w=2$) is discussed in SI.

\subsection{Compact (recurrent) case} 

\begin{figure}[h!]
    \centering
    \includegraphics[scale=0.78]{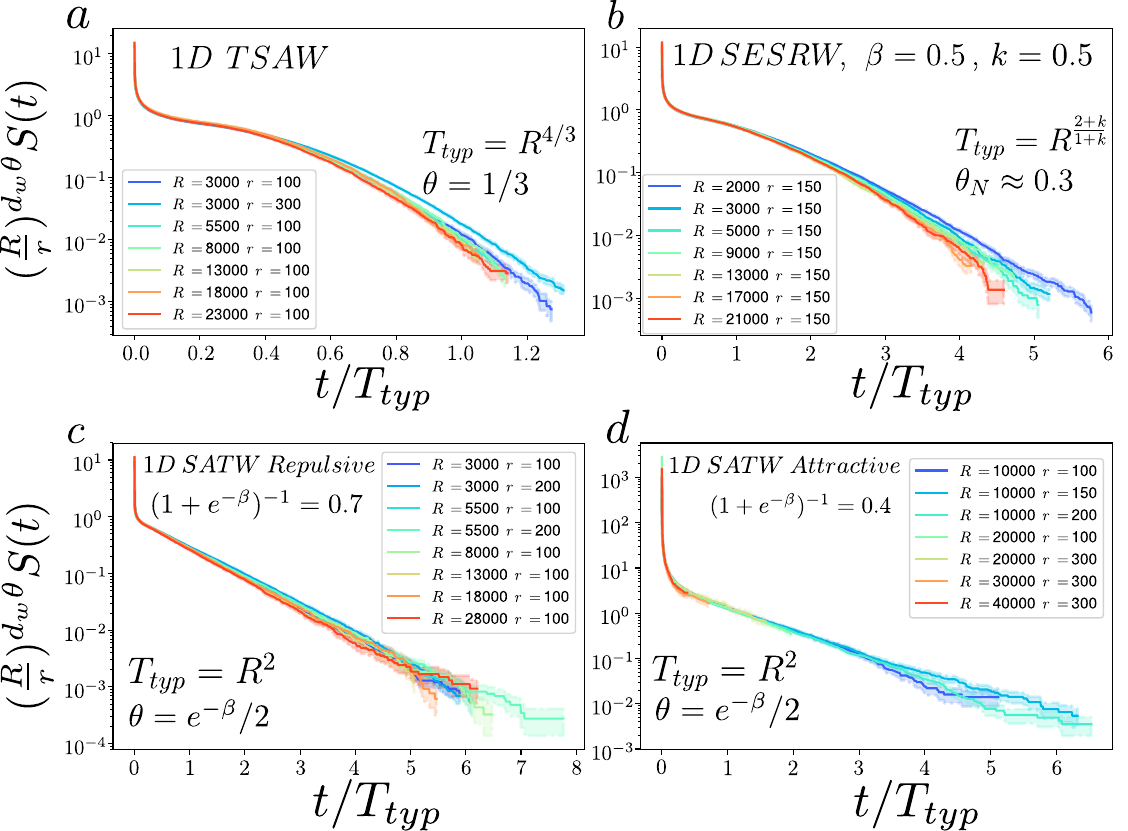}
    \caption{Asymptotic  FPT distribution of compact self-interacting random walks in confined domains for fresh initial conditions. Here $S(t)$ is the survival probability of the random walker,  whose scaling with geometrical parameters is deduced from \eqref{Fbar}. The collapse of numerical simulations after rescaling for different values of geometric parameters captures the dependence of the FPT distribution on the geometric parameters $r,R$. Simulations are performed in $1D$ boxes of size $R$ with reflecting boundary conditions. \textbf{a.} $1D$ TSAW with $d_w = 3/2$ (independent of $\beta$). \textbf{b.} $1d$ SESRW with $\beta=0.5$ and $k=0.5$. $1d$ repulsive SATW  with $(1 + e^{-\beta})^{-1} = 0.7$ ( \textbf{c.}) and $(1 + e^{-\beta})^{-1} = 0.4$ (\textbf{d.}).}
    \label{fig5}
\end{figure}

We sketch in this section the derivation of the asymptotic  FPT distribution $F(t,r,R)$  for compact self-interacting random walks  in the large volume limit $R\to \infty$. For compact processes the FPT distribution is independent of the target linear size $a$ for $r\gg a$ ; we focus on this regime below. Following \cite{Levernier:2018qf}, $F(t,r,R)$ can be written  as a partition over trajectories that either hit  the reflecting boundary before the target (with probability $\pi(r,R)$ and conditional FPT distribution to the target  $F_b(t,r,R)$) or hit the target before the boundary (with probability $\pi(r,R)$ and conditional FPT distribution to the target  $F_t(t,r,R)$):
\begin{equation}\label{part}
F(t,r,R) = \pi F_b(t,r,R) + (1-\pi)F_t(t,r,R).
\end{equation} 
Importantly, the weight $1-\pi$ of trajectories that hit the target first can  be expressed in the limit $R\to \infty$ (with $r$ fixed) in terms of the FPT distribution in infinite space  $F_\infty(t,r)$ : 
\begin{equation}\label{pi1}
 \pi(r,R) \underset{R\gg r}{\propto} \int^\infty_{R^{d_w}} F_\infty (t,r) dt,
\end{equation} 
which expresses the fact that most trajectories that hit the target before the boundary yield a FPT smaller than the timescale $R^{d_w}$.    Making use of the definition of $\theta$ for processes in infinite space, we then obtain from dimensional analysis
\begin{equation}
 F_\infty(t,r)\propto \frac{r^{d_w\theta}}{t^{\theta +1}}
\end{equation} 
in the regime $1\ll t\ll R^{d_w}$, which yields from \eqref{pi1}:
\begin{equation}\label{pi2}
\pi(r,R) \underset{R\gg r}{\propto} \left(\frac{r}{R}\right)^{d_w\theta}.
\end{equation} 
We stress that here the persistence exponent $\theta$ is defined in infinite space, and not in confined geometry. Next, the above argument leading to \eqref{pi1} also implies that 
\begin{equation}\label{Ft}
 F_t(t,r,R)\propto \Theta(t/R^{d_w}) F_\infty(t,r)\propto \Theta(t/R^{d_w}) \frac{r^{d_w\theta}}{t^{\theta +1}},
\end{equation} 
where $\Theta$ denotes a step function with $\Theta(x\ll1)=1$ and $\Theta(x\gg1)=0$.   At this stage, the conditional FPT distribution $F_b(t,r,R)$ remains to be determined. By definition, this quantity involves trajectories that interact with the domain boundary. However, our analysis above shows that the increments of compact processes are identical in confined and infinite geometries. In the limit $R\to \infty$ with $r$ fixed, $F_b(t,r,R)$ can thus depend  only on the time scales  $t$ and $R^{d_w}$; dimensional analysis then  yields the following scaling form:
\begin{equation}\label{Fb}
 F_b(t,r,R)\sim g(t/R^{d_w})/t,
\end{equation} 
where $g$ is an undetermined function that depends on the process. Finally, it is convenient to introduce the rescaled variable $\eta=t/R^{d_w}$, and write, from \eqref{part},\eqref{pi2},\eqref{Ft} ,\eqref{Fb} its asymptotic distribution for $R\to \infty$ for $\eta>0$ with $r$ fixed :
\begin{equation}\label{Fbar}
 {\bar F}(\eta,r,R)= \left(\frac{r}{R}\right)^{d_w\theta} h(\eta)
\end{equation} 
where $h$ is an undetermined function that depends on the process. Finally, this explicitly captures  the dependence of the FPT distribution on the geometrical parameters $r,R$, and therefore of all its moments (when they exist). In particular, the mean FPT can be readily derived and satisfies:
\begin{equation}\label{MFPTcompact}
 \langle T\rangle\propto R^{d_w(1-\theta)}  r^{d_w\theta}.
\end{equation} 
The mean FPT thus scales non linearly with the confining volume $V\sim R^{d}$ (because one has $\theta\not=1-d/d_w$) for  SESRW and SATW, as was found for other examples of aging processes ; notably, this scaling is linear for the TSAW. Strikingly, the asymptotic form of the FPT distribution \eqref{Fbar} is comparable to that obtained in \cite{Levernier:2018qf}, and  can be determined solely from the knowledge of $d_w,\theta$, which are defined in infinite space. This holds even if the dynamics  of the process  is ultimately impacted by the geometric confinement, as we have shown above -- this result is in particular independent of the persistence exponent in confinement $\theta_c$.  Fig.\ref{fig5} shows an  excellent quantitative agreement between numerical simulations and this  analytical result. The data collapse of the properly rescaled FPT distribution shows that our approach fully  captures its  dependence on both $r$ and $R$ for all examples of compact self-interacting random walks that we have studied.

\subsection{Non compact (transient) case} 

\begin{figure}[h!]
    \centering
    \includegraphics[scale=0.55]{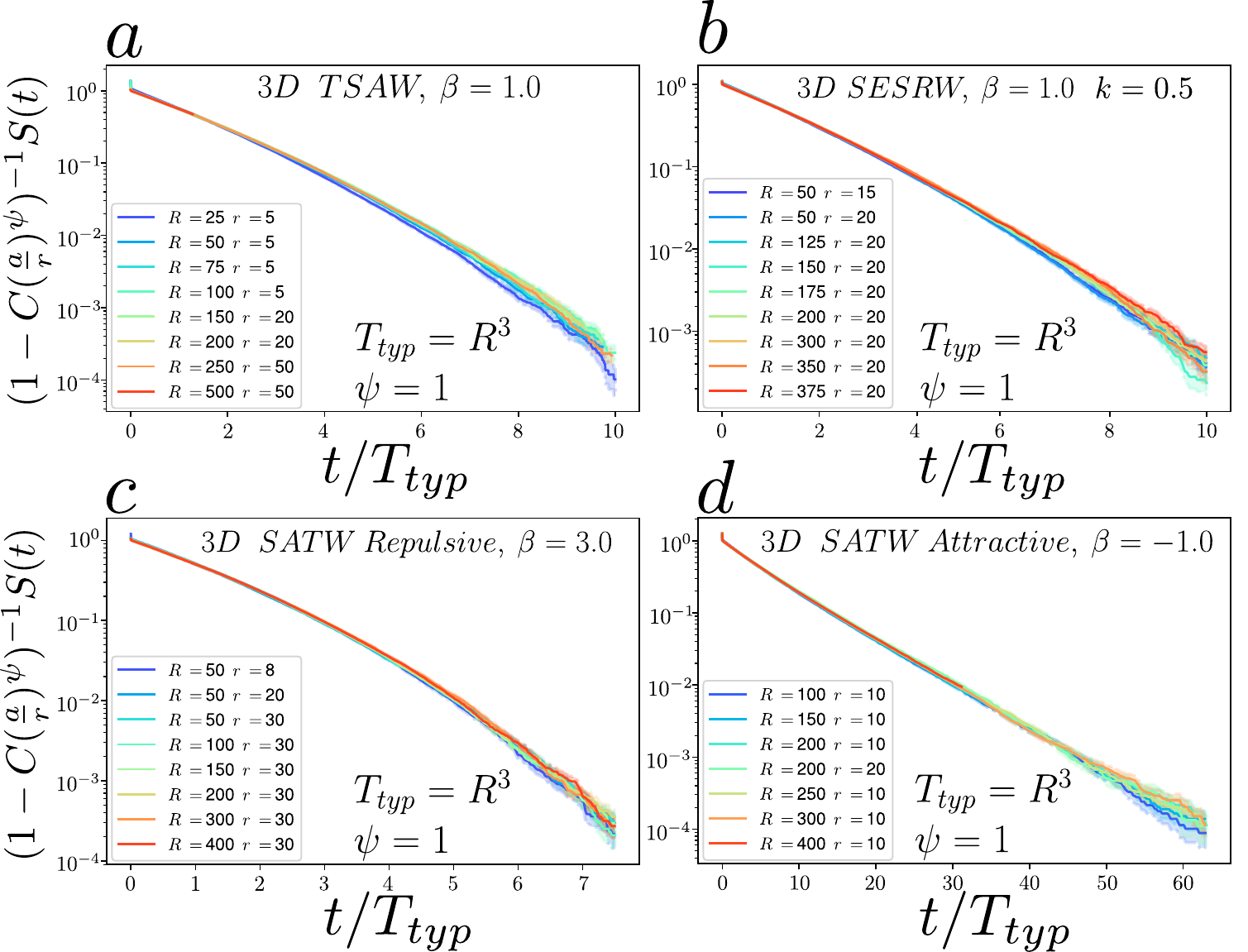}
    \caption{Asymptotic  FPT distribution of non compact self-interacting random walks in confined domains for fresh initial conditions. Here $S(t)$ is the survival probability of the random walker,  whose scaling with geometrical parameters is deduced from \eqref{Fbar2}. The collapse of numerical simulations after rescaling for different values of the geometrical parameters captures the dependence of the FPT distribution on $r,R$. Simulations are performed in $3d$ boxes of size $R$ with reflecting boundary conditions and the constant $C$ is measured numerically. \textbf{a.} $3d$ TSAW with $\beta=1.0$. \textbf{b.} $3d$  SESRW with $\beta=1.0$ and $k=0.5$. $3d$ SATW repulsive with $\beta=3.0$ (\textbf{c}) and with $\beta = -1.0$ (\textbf{d}). }
    \label{fig6}
\end{figure}

We now turn to the non-compact case. As opposed to the compact case, in the regime $r\gg a$ that we consider below, the FPT distribution depends on $a$.  Following \cite{Levernier:2018qf}, we  call  excursion a fraction of trajectory that starts  from the sphere $S$ of radius $R/2$ centered on the target, next hits  the boundary and eventually returns to $S$. The FPT distribution can then be written as a partition over the number $n$ of excursions  before the first-passage to the target, where we introduce $\Phi_n(t)$ as the corresponding conditional FPT distribution : 
\begin{equation}
%F(t,a,r,R)=p_0 \frac{1}{t}\phi_0\left(\frac{t}{r^{d_w}}\right) + \sum_{n=1}^\infty \frac{1}{t}\phi_n\left(\frac{t}{t_n}\right)(1-p_0)(1-p)^{n-1}p,
F(t,a,r,R)=p_0 \Phi_0(t) + (1-p_0)\sum_{n=1}^\infty \Phi_n(t)P(n). 
\label{nc1}
\end{equation}
Here $p_0\sim (a/r)^\psi$  is the probability to hit the target before the boundary  starting from $r$, and $P(n)$ the probability that the target is reached for the first time during the $n^{th}$  excursion. This can be written
\begin{equation}
P(n)=p_n\prod_{k=1}^{n-1} (1-p_k)
\label{Pn}
\end{equation}
where $p_k$ is the probability that the target is found during the $k^{th}$  excursion, knowing that is has not been found before.  Our  analysis of increments and transience exponents $\psi,\psi_c$ above (see \eqref{incnc}) shows that,  in confinement,  non compact self-interacting random walks are diffusive for $t\gg 1$ in both regimes $T\ll R^{d_w}$ (with diffusion coefficient $D_<$) and $T\gg R^{d_w}$  (with diffusion coefficient $D_>$). We thus denote by $D_n\equiv D_>+\delta D_n$ the effective diffusion coefficient during the  $n^{th}$  excursion, which verifies $|\delta D_n|\le |D_<- D_>|$ and $\delta D_n\to 0$ for $n\gg 1$.  In addition, one has $\psi_c=\psi$. We can thus write   $p_k\sim (C_> +\delta C_k) (a/R)^\psi$, where $\delta C_k\to 0 $ for $k\gg 1$.  Note that here we have implicitly assumed (and checked numerically, see SI) that the conditional probability $p_k$ behaves as the unconditional probability  that the target is found during the $k^{th}$ excursion.
%where $p\sim (a/R)^\psi$  is   the probability to reach the target before the boundary starting from $S$  and . Note that here we implicitly assume that excursions are independent in the large $R$ limit. Physically, it originates from  the divergence with $R$ of the typical time $\tau_n$ needed to perform the $n^{th}$ excursion, which hence can be taken larger than  all correlation times of the process,  as was checked numerically (see appendix \ref{detailednoncompact}). 
Last, 
a  scaling argument   (see  SI) shows that 
\begin{equation}
\Phi_n(t)=\frac{1}{t}\phi(t/t_n), 
\end{equation}
where $t_n$ is the typical time elapsed before the $n^{th}$  excursion, which verifies
\begin{equation}
t_n=R^{d_w}\sum_{k=1}^{n-1}\frac{1}{D_k}.
\end{equation}
Finally, taking the $R\to\infty $ limit in \eqref{nc1} with $r,a$ fixed, one finds that the rescaled variable $\eta=t/R^d$ admits asymptotically the following distribution for $\eta\not=0$ (see SI):
\begin{equation}\label{Fbar2}
 {\bar F}(\eta,r,R)= \left(1-C\left(\frac{a}{R}\right)^{\psi}\right)  h(\eta)
\end{equation} 
where $h$ is an undetermined process dependent scaling function -- not necessarily exponential --  and $C$ a process dependent constant. Similarly to the compact case, this explicitly captures  the dependence of the FPT distribution on the geometrical parameters $r,R$, and therefore of all its moments (when they exist). In particular, the mean FPT is given by :
\begin{equation}\label{MFPTnoncompact}
 \langle T\rangle\sim \frac{R^{d}}{a^{\psi}} \left(1-C\left(\frac{a}{R}\right)^{\psi}\right).
\end{equation} 
In contrast to the compact case, the mean FPT thus scales  linearly with the confining volume $V\sim R^{d}$. Remarkably, the asymptotic form of the FPT distribution \eqref{Fbar} is comparable to that obtained in \cite{Levernier:2018qf} in absence of power-law aging, and  can be determined solely from the knowledge of $d_w,\psi$, which are defined in infinite space. This holds even if the dynamics  of the process  is  impacted by the geometric confinement, as we have shown above. However, geometric confinement does not change the diffusive scaling of non compact self-interacting walks ; \eqref{Fbar2} shows that this is sufficient to preserve the dependence on $r,R$ of  the FPT distribution  in confinement.    Fig.\ref{fig6} shows an  excellent quantitative agreement between numerical simulations and this  analytical result. The data collapse of the properly rescaled FPT distribution shows that our approach fully  captures its  dependence on both $r$ and $R$ for all examples of non compact self-interacting random walks that we have studied.

\section{Aged initial conditions }
In this section, we analyse the impact of initial conditions on the FPT statistics of confined self-interacting random walks. As we have shown above, the dynamics of self-interacting random walks display aging properties, which can depend on geometric confinement. In other words, the dynamics is different if the random walk starts at $T=0$ (fresh initial conditions studied above, for which the number of visits to any site $i$ of the domain is set to $n_i=0$)  or at $T\gg R^{d_w}$ (aged initial conditions, for which $n_i\gg 1$). We show that the FPT distribution can be readily obtained for aged initial conditions  by adapting the approach developed above for fresh initial conditions, and highlight the impact of initial conditions.

\subsection{Compact (recurrent) case} 
\begin{figure*}
   % \centering
    \includegraphics[scale=0.95]{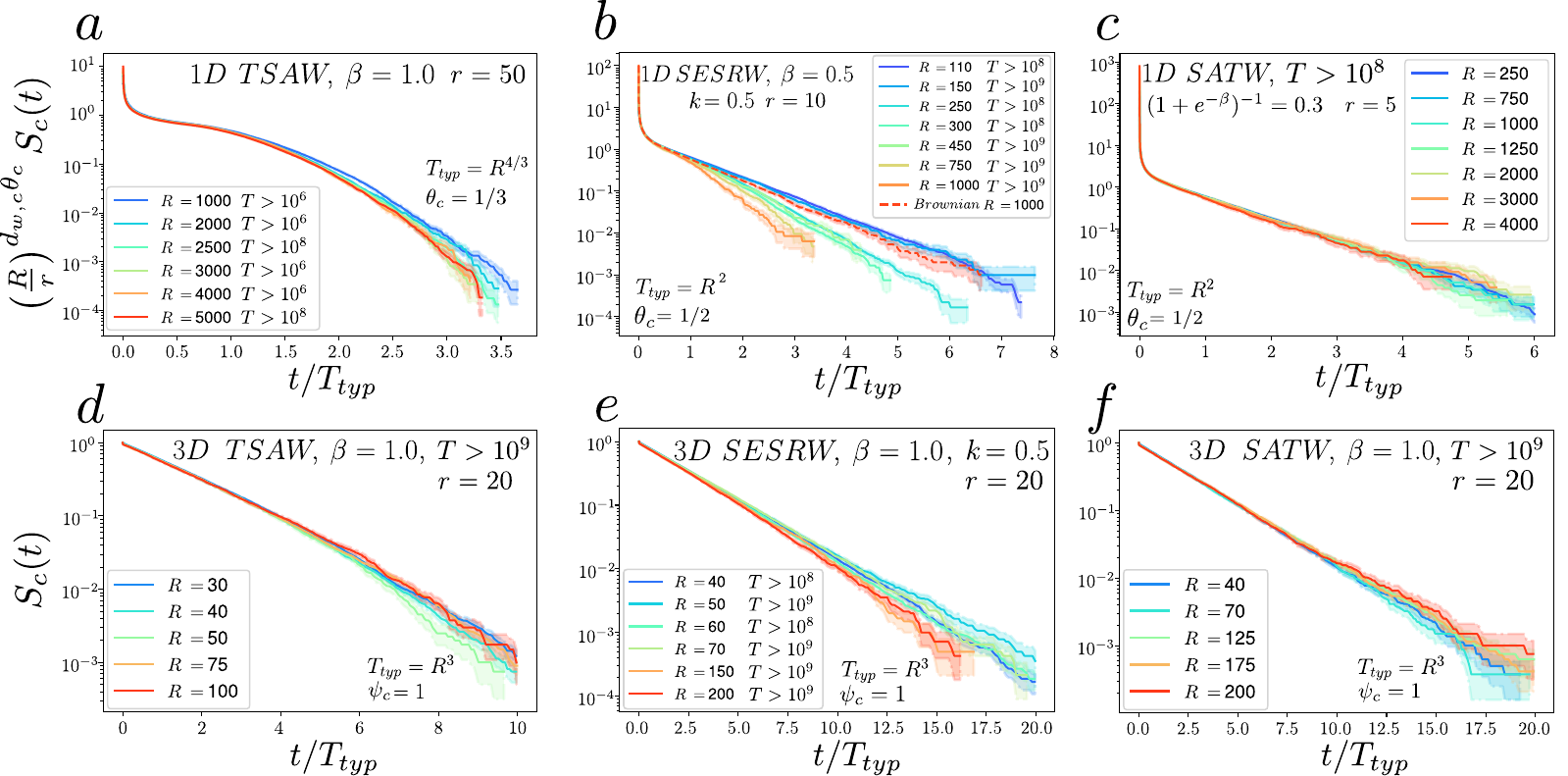}
    \caption{Asymptotic  FPT distribution of  self-interacting random walks in confined domains  for aged initial conditions. The search process starts at $t=0$, but the random walker is assumed to have explored the domain from $t=-T$ to $t=0$.  $S(t)$ is the survival probability of the random walker at time $t$. The scaling of $S(t)$ with geometrical parameters is deduced from \eqref{Fbarc} and \eqref{Fbar2c}, for compact and non-compact processes respectively. The collapse of numerical simulations after rescaling  captures the dependence of the FPT distribution on geometrical parameters. Simulations are performed in $1d$ and  $3d$ boxes of size $R$ with reflecting boundary conditions for  fixed $r$ and $a$. Compact cases : \textbf{a.} $1d$ TSAW with $\beta=1.0$; \textbf{b.} $1d$ SESRW with $\beta=1.0$ and $k=0.5$. The FPT distribution of the  simple random walk  is added for comparison (dashed curve); \textbf{c}. $1d$ SATW repulsive with $\beta=3.0$  Non-compact cases : \textbf{d}. $3d$ TSAW with $\beta=1.0$; \textbf{e}. $3d$ SESRW with $\beta=1.0$ and $k=0.5$; \textbf{f}. $3d$ SATW with $\beta=1.0$.}
    \label{fig7}
\end{figure*}

For aged initial conditions, because $T\gg R^{d_w}$, the only relevant regime is $t\ll T$. In this regime we have found the following behaviour of the increments:
\begin{equation}
\Delta^2(T,t)\sim 2 D_<(t) t^{2/d_w}\propto t^{2/d_{w,c}}
\end{equation}
where the effective walk dimension $d_{w,c}$ can be different from $d_w$ (see  SESRW in Fig.\ref{fig3}). The relevant persistent exponent is clearly $\theta_c$ in this regime. All steps leading to the derivation of the FPT distribution (see previous section) can then be reproduced. It is found that 
 the rescaled variable $\eta=t/R^{d_{w,c}}$ is  asymptotically  distributed according to :
\begin{equation}\label{Fbarc}
 {\bar F}_c(\eta,r,R)= \left(\frac{r}{R}\right)^{d_{w,c}\theta_c} h_c(\eta)
\end{equation} 
where $h_c$ is an undetermined function that depends on the process. Initial conditions can thus deeply impact the FPT distribution, and even its scaling form : they can modify the walk dimension $d_{w,c}$, the persistence exponent $\theta_c$, and the scaling function $h_c$. This result is confirmed by numerical simulations (see Fig.\ref{fig7}). For the SATW, one has $\theta_c\not=\theta$ and $d_{w,c}=d_w=2$ while  for the SESRW one has $\theta_c\not=\theta$ and $d_{w,c}\not=d_w$ ; the scaling of the FPT distribution is thus modified by initial conditions for these processes. In contrast, for the TSAW one has $\theta_c=\theta$ and $d_{w,c}=d_w$ and the  scaling of the FPT distribution is not  modified by initial conditions.

\subsection{Non compact (transient) case} 
In the regime $T\gg R^{d_w}$ and $1\ll t\ll T$, we have found the following diffusive scaling of  increments for non compact self-interacting random walks:
\begin{equation}
\Delta^2(T,t)\underset{t\gg 1}{\sim} 2 D_> t. 
\end{equation}
In addition, we have shown that $\psi_c=\psi$. All steps leading to the derivation of the FPT distribution (see previous section) can then be straightforwardly reproduced. It is found that 
\begin{equation}\label{Fbar2c}
 {\bar F}_c(\eta,r,R)= \left(1-C\left(\frac{a}{R}\right)^{\psi}\right)  h_c(\eta)
\end{equation} 
where $\eta=t/R^d$ and $h_c$ is an undetermined function that depends on the process. In the case of non compact self-interacting random walks, initial conditions thus do not modify the scaling of the FPT distribution ; they however can change the scaling function $h_c$. This result is confirmed by numerical simulations (see Fig.\ref{fig7}).

\section{Discussion and conclusion}

\subsection{Summary of the results}
Our joint analytical and numerical analysis shows finally that long range memory effects can have deep consequences on the dynamics of generic self-interacting random walks ; they can induce  aging (quantified by $D(T,t)$ and $d_w$) and non trivial persistence and transience exponents $\theta$ and  $\psi$, which we characterized quantitatively. In striking contrast with other non Markovian processes, we have shown that geometric confinement  can strongly  modify   the  dynamic properties  of self-interacting random walks, beyond imposing locally reflecting boundary conditions : the dynamics of increments can be modified (in the non compact case), as well as  persistent exponents (in the compact case).

Based on this systematic quantitative analysis, we have  shown that universal scaling forms of the FPT distributions of general self-interacting random walks in confinement  can be derived in the large volume limit, by generalizing the approach introduced in \citep{Levernier:2018qf}.   For "fresh" initial conditions,   we find that the   FPT distribution in confinement can be asymptotically predicted from the knowledge of the process in infinite space only (via the infinite space exponents $d_w, \theta, \psi$) : geometric confinement ultimately does modify the dynamics of the process and even changes  the corresponding   exponents $d_{w,c}, \theta_c, \psi_c$  defined in confinement, but this occurs only at timescales larger than the typical FPT, and thus only mildly impacts the FPT statistics.
In contrast, for aged initial conditions the  exponents that determine the FPT distribution are those defined in confinement $d_{w,c}, \theta_c, \psi_c$ , and can thus be different from the classical infinite space exponents $d_w, \theta, \psi$. In that case, the process in confinement must therefore be characterized to determine the FPT distribution.

\subsection{Search efficiency of self-interacting random walkers}
These results allow us to assess  the efficiency of space exploration of self-interacting random walks, and in particular to discuss the impact of memory effects on target search kinetics.

In infinite space, attractive self-interactions ($\beta<0$) can have drastic consequences on space exploration :    for bounded effective interaction potentials $V(n)$ (SATW), the random walk  is subdiffusive for $d=2,3$ and $|\beta|>|\beta|_c$ and characterised by $\theta=\infty$, so that all moments of the FPT to a target  are finite. In this case memory effects thus give a decisive advantage to attractive SATW ($|\beta|>|\beta|_c$) as compared to normal random walks ($\beta=0$) or repulsive self-interacting walks ($\beta>0$).

 In confined domains, the discussion is very different. If no prior information on the target position is available, the relevant observable to quantify the search kinetics is the position averaged mean FPT ${\bar{\langle T\rangle}}$. For compact processes, our results yield ${\bar{\langle T\rangle}}\propto R^{d_w}$. Search kinetics is thus enhanced by lowering $d_w$, which amounts to maximising  the scaling of the MSD with time. In that case, memory effects  give a decisive advantage to repulsive $1d$ TSAW and $1d$ SESRW, which both show a superdiffusive exponent $d_w< 2$ for all values of $\beta>0$.  For non compact processes, we obtained ${\bar{\langle T\rangle}}\propto R^{d}$, which is consistent with the   large time diffusive limit of non compact  (repulsive or attractive) self-interacting random walks. The scaling of ${\bar{\langle T\rangle}}$ with $R$ is thus independent of memory effects, which however modify the effective diffusion coefficient and are thus favorable in the repulsive case.

 If the starting distance $r$ from the target is known, the full FPT distribution is needed to analyse the search kinetics. For compact processes, our results \eqref{Fbar} show that the set of trajectories that hit the target  can be decomposed into a set of fast trajectories, with timescale $\propto r^{d_w}$ and weight $1-\alpha (r/R)^{d_w \theta}$ (where $\alpha$ is a constant), and a set of slow trajectories that typically hit the domain boundaries before the target, with timescale $\propto R^{d_w}$ and weight $\propto (r/R)^{d_w \theta}$. The exponents $d_w,\theta$ thus appear as key parameters that control the respective weight of fast and slow trajectories, as well as the typical timescale of slow trajectories. For random processes with stationary increments, it has been proposed that both exponents are not independent and satisfy $\theta=1-d/d_w$ \cite{Bray:2013,Meroz:2011tu,Levernier:2018qf} ; in that case, increasing the weight of direct, fast trajectories by increasing $d_w$ comes at the cost of increasing the timescale of indirect trajectories. This is also the case of the $1d$ repulsive TSAW, for which we found numerically $\theta=1-d/d_w$. In the case of the $1d$ SESRW and the $1d$ SATW however, we found that $d_w$ and $\theta $ are independent, with a dependence of $\theta $ only on the coupling parameter $\beta$. This shows that repulsive self-interactions can be favorable for large starting distances because they diminish the timescale of indirect trajectories by lowering $d_w$ ($1d$ TSAW and $1d$ SESRW) ; they however in all cases reduce the weight of direct trajectories and are thus detrimental at short distances. In turn, attractive interactions (SATW) are detrimental for $d=2,3$ because they increase the timescale of indirect trajectories by increasing $d_w$ (subdiffusive SATW), while they preserve the diffusive scaling for $d=1$ ; they can however significantly increase the  weight of direct trajectories by increasing $\theta$ ($d=1,2,3$), and are thus favorable at short distances. Finally, in the non compact case, our results \eqref{Fbar2} show that  the FPT statistics is characterized by the single time scale $R^d$ as long as $r\gg a$. As in the case of the position averaged mean FPT,  memory effects modify only the effective diffusion coefficient ; they are thus favorable in the repulsive case, but do not impact scaling properties of the FPT distribution.

 Finally, this analysis shows that memory effects induced self-interactions can have a deep impact on space exploration, as quantified by various observables. Qualitatively, attractive self-interactions have dramatic effects and can lead to subdiffusion with compact exploration, which is favorable for local exploration,  and even to  self-trapping. Repulsive self-interactions have important effects for compact random walks, for which they modify the walk dimension $d_w$ and thus the scaling of the position averaged mean FPT with the size of the confining domain ; this  is thus favorable for global exploration of confined domains with no prior informations on the target position.

%\bibliographystyle{apsrev4-2} 
%\bibliographystyle{vincent} 
%\bibliography{url} 
%\bibliography{../../biblio/biblinew.bib,ERW-SATW.bib}
%\bibliography{BibFused.bib}
%\bibliography{../../biblio/biblinew.bib,BibFused.bib}


\end{document}